\documentstyle[12pt]{article}
\def\ap#1#2#3{Ann.\ Phys.\ (NY) #1 (19#3) #2}

\def\jmp#1#2#3{J.\ Math.\ Phys.\ #1 (19#3) #2}

\def\np#1#2#3{Nucl.\ Phys.\ B#1 (19#3) #2}
\def\pl#1#2#3{Phys.\ Lett.\ #1B (19#3) #2}
\def\pr#1#2#3{Phys.\ Rev.\ D #1 (19#3) #2}
\def\prb#1#2#3{Phys.\ Rev.\ B #1 (19#3) #2}
\def\prep#1#2#3{Phys.\ Rep.\ #1 (19#3) #2}

\def\rmp#1#2#3{Rev.\ Mod.\ Phys.\ #1 (19#3) #2}

\def\cmp#1#2#3{Comm.\ Math.\ Phys.\ #1 (19#3) #2}
\def\cmp#1#2#3{Comm.\ Math.\ Phys.\ #1 (19#3) #2}
\def\nc#1#2#3{Il Nuovo Cimento #1A (19#3) #2}
\def\lnc#1#2#3{Lettere al Nuovo Cimento #1 (19#3) #2}

\newskip\humongous \humongous=0pt plus 1000pt minus 1000pt
\def\caja{\mathsurround=0pt}
\def\eqalign#1{\,\vcenter{\openup1\jot \caja
        \ialign{\strut \hfil$\displaystyle{##}$&$
        \displaystyle{{}##}$\hfil\crcr#1\crcr}}\,}
\newif\ifdtup

\def\frac#1#2{ {{#1} \over {#2} }}
\def\ie{\hbox{\it i.e.}{ }}

\def\half{\mbox{\small $\frac{1}{2}$}}

\def\re#1{(\ref{#1})}
\def\beq{\begin{equation}}
\def\eeq{\end{equation}}
\def\beeq{\begin{eqnarray}}
\def\beeqn{\begin{eqnarray*}}
\def\eeeq{\end{eqnarray}}
\def\eeeqn{\end{eqnarray*}}
\def\se{S_{\mbox{\footnotesize{eff}}}}
\def\ser{S_{\mbox{\footnotesize{eff,rel}}}}
\def\sei{S_{\mbox{\footnotesize{eff,irr}}}}
\def\sbrs{S_{\mbox{\scriptsize{BRS}}}}
\def\set{S_{\mbox{\scriptsize{eff}}}}

\def\Gr{\G_{\mbox{\footnotesize{rel}}}}
\def\Grt{\tilde \G_{\mbox{\footnotesize{rel}}}}

\def\De{\D_{\mbox{\footnotesize{eff}}}}
\def\Dee{\D_{\mbox{\footnotesize{eff}},2}}
\def\Den{\D_{\mbox{\footnotesize{eff}},n}}
\def\Denn{\D_{\mbox{\footnotesize{eff}},n'}}

\def\Dei{\D_{\mbox{\footnotesize{eff,irr}}}}
\def\DG{\D_{\G}}
\def\DGr{\D_{\G,\mbox{\footnotesize{rel}}}}

\def\bc{\bar c}

\def\r{\rho}
\def\br{\bar \rho}
\def\de{\delta}

\def\G{\Gamma}

\def\L{\Lambda}

\def\g{\gamma}
\def\D{\Delta}
\def\dL{\L\partial_\L}
\def\UV{$\L_0\to\infty\;$}
\def\IR{$\L\to 0\;$}
\def\bit{\begin{itemize}}
\def\eit{\end{itemize}}
\def\ben{\begin{enumerate}}
\def\een{\end{enumerate}}

\def\nome#1{{\label{#1}}}

\begin{document}
\begin{titlepage}
\renewcommand{\thefootnote}{\fnsymbol{footnote}}
\begin{flushright}
     UPRF 94-414\\
     November 1994 \\
\end{flushright}
\par \vskip 10mm
\begin{center}
{\Large \bf
BRS symmetry\\
from renormalization group flow}
\end{center}
\par \vskip 2mm
\begin{center}
        {\bf M.\ Bonini, M.\ D'Attanasio} \\
        Dipartimento di Fisica, Universit\`a di Parma and\\
        INFN, Gruppo Collegato di Parma, Italy\\
        and\\
        {\bf G.\ Marchesini}\\
        Dipartimento di Fisica, Universit\`a di Milano and\\
        INFN, Sezione di Milano
\end{center}
\par \vskip 2mm
\begin{center} {\large \bf Abstract} \end{center}
\begin{quote}
By using the exact renormalization group  formulation we prove
perturbatively the Slavnov-Taylor (ST) identities in $SU(2)$
Yang-Mills theory. This results from two properties: {\it locality},
\ie the ST identities are valid if their local part is valid;
{\it solvability}, \ie the local part of ST identities is valid if
the couplings of the effective action with non-negative dimensions
are properly chosen.
\end{quote}
\end{titlepage}

\noindent
{\bf{1. Introduction}}
\vskip 0.3cm
\noindent
A renormalized theory is defined by giving the ``relevant part'' of the
effective action, \ie its local part involving couplings which have
non-negative mass dimension\footnote{In this paper the relevant
parts include also what is usually called marginal.}.
If the fields have non-zero mass these relevant couplings
could be given
by the first coefficients of the Taylor expansion of vertex functions
around zero momenta.
If the fields have zero mass the expansion must be done around some
non-vanishing Euclidean subtraction point $\mu \ne 0$.
For the massless $\Phi^4_4$ theory there are three relevant couplings
corresponding to the physical mass, wave function normalization
and interaction strength $g$ at the subtraction point $\mu$.
These three physical couplings define completely the theory.

In a gauge theory the effective action contains more couplings than
physical parameters.
In the $SU(2)$ Yang-Mills theory for instance the effective action
$\G[\phi,\g]$ (with $\phi=(A_\mu,c)$ the vector and ghost fields
and $\g=(u_\mu,v)$ the BRS sources)
contains nine relevant couplings but only three are fixed by the
vector and ghost field normalizations and by the interaction strength
$g$ at a subtraction point $\mu$.
After fixing these three physical parameters, the relevant part of
the Yang-Mills effective action can be written
$$
\Gr[\phi,\g] \equiv T_4^{(\mu)}\,\G[\phi,\g]
=\sbrs[\phi,\g]
+\hbar\Grt[\phi,\g;\r_i]\,,
$$
where
we denote by $T_d^{(\mu)}$ the operator which extracts from a
given functional of dimension $d-4$ its relevant part (in four
space-time dimension) with a non-vanishing subtraction point $\mu$.
For $SU(2)$ the BRS classical action \cite{BRS}, in the Feynman gauge,
is given by
$$
\sbrs=\int d^4x \biggl\{-\frac 1 4 F_{\mu\nu}^2
-\frac 1 2 (\partial_\mu A_\mu)^2
+ \frac 1 g w_\mu \cdot D_\mu c - \frac 1 2 \, v\cdot c \wedge c
\biggr\}\,
$$
and
\beq\nome{tG}
\eqalign{
\Grt[\phi,\g;\r_i]
&
\equiv
\int d^4x \biggl\{
\r_1 \, \frac 1 2 A_\mu^2
+\r_2 \,\frac 1 2 (\partial_\mu A_\mu)^2
+\r_3\, w_\mu \cdot c\wedge A_\mu
+\r_4 \, \frac 1 2 v\cdot c \wedge c
\cr&
+\r_5\, \frac{g^2}{4}(A_\mu\wedge A_\nu)^2
+\r_6\, \frac{g^2}{4} (A_\mu\cdot A_\nu)^2
\biggr\}\,
\,,
}
\eeq
with
$F_{\mu\nu}=\partial_\mu A_\nu- \partial_\nu A_\mu +
g A_\mu \wedge A_\nu$, $D_\mu c =\partial_\mu c +gA_\mu \wedge c$
and $w_\mu=u_\mu+g \partial_\mu \bc$.
The six couplings $\r_i$ in $\Grt$ vanish at tree level and
should be constrained by the gauge symmetry.
For instance $\r_1$ is the vector field mass and we expect it must
vanish.
The gauge symmetry requires that the effective action satisfies the ST
identities
\beq\nome{ST}
\D_\G[\phi,\g] \equiv {\cal S}_{\G'}\, \G'[\phi,\g]=0\,,
\;\;\;\;\;\;
\;\;\;\;\;\;
\G'[\phi,\g]=\G[\phi,\g]+
\frac 1 2 \int d^4x (\partial_\mu A_\mu)^2\,,
\eeq
where
${\cal S}_{\G'}$ is the usual Slavnov operator (see for instance
\cite{B1}).
Also for $\D_\G[\phi,\g]$ one defines the relevant part.
Since this functional has dimension one we have
$$
\DGr[\phi,\g;\de_i]=
T_5^{(\mu)} \D_\G[\phi,\g]
\,,
$$
where $\de_i$ are the relevant parameters, \ie the coefficients
of monomials in the fields, sources and momenta of dimension not
greater than five.
For the $SU(2)$ case, there are eleven relevant parameters $\de_i$
and we have
\newpage
$$
\DGr[\phi,\g;\de_i]=\int d^4x \, \biggl\{
\de_1 \,A_\mu\cdot\partial_\mu c
-\de_2 \,A_\mu\cdot\partial^2\partial_\mu c
+\de_3 \,A_\mu\cdot (\partial^2 A_\mu)\wedge c
+\de_4 \,A_\mu\cdot (\partial_\mu \partial_\nu A_\nu)\wedge c
$$
$$
+\half \de_5 \,(\partial_\mu w_\mu)\cdot c\wedge c
+\half \de_6 \,(w_\mu \wedge A_\mu) \cdot (c\wedge c)
+\de_7 \, ((\partial_\mu A_\mu) \cdot A_\nu) (A_\nu \cdot c)
+\de_8 \, ((\partial_\mu A_\mu) \cdot c) (A_\nu \cdot A_\nu)
$$
\beq\nome{DGrel}
+\de_9 \, ((\partial_\nu A_\mu) \cdot A_\nu) (A_\mu \cdot c)
+\de_{10} \, ((\partial_\nu A_\mu) \cdot A_\mu) (A_\nu \cdot c)
+\de_{11} \, ((\partial_\nu A_\mu) \cdot c) (A_\mu \cdot A_\nu)
\biggr\} \,.
\eeq
The important question is then whether it is possible to fix the
couplings $\r_i$ in such a way to ensure that the full set of
ST identities \re{ST} is satisfied.

In perturbation theory this question is solved in the positive sense
since almost two decades.
In dimensional regularization with minimal subtraction
\cite{tHV}-\cite{BM}
of gauge theories without chiral fermions both the bare Lagrangian
and the regularization procedure do not break the
BRS symmetry so that the effective action (and then also the
couplings $\r_i$) automatically satisfies ST identities.
In chiral gauge theories dimensional regularization breaks the symmetry
\cite{tHV,BM,gamma5}.
However, if no anomalies are present, one can implement the ST
identities by introducing non-invariant local counterterms \cite{chiral}.
This is done by a so called ``fine tuning procedure''.
This fact is independent of the regularization since the classification
of all possible anomalies is a purely algebraic problem
\cite{B1,cohom}, \ie anomalies are associated to the existence of
non-trivial cohomology classes of the Slavnov operator.

Becchi has recently \cite{B} shown that exact renormalization
group (RG) flow \cite{W}-\cite{G} can be used to deduce the ST
identities.
By taking advantage of the fact that in the RG flow one introduces
an infrared (IR) cutoff $\L$, he has defined the relevant couplings at
the non-physical point $\L \ne 0$. This allowed him to use a
vanishing subtraction point $\mu=0$.
The connection with the physical couplings defined in \re{tG} is then
indirect but can be obtained perturbatively.
The ST identities are then analyzed at $\L \ne 0$ and then, because of
the IR problem, one cannot continue to the physical point $\L=0$.

In this note we follow the same analysis but work at the physical
point $\L=0$ and prove directly the ST identities \re{ST} for the
effective action $\G[\phi,\g]$.
We are then able to discuss the symmetry in terms of the relevant
couplings defined in \re{tG}.
In a more detailed paper \cite{BDM} there will be the full calculations
for the $SU(2)$ case and the precise expression of the
values of the six couplings $\r_i$.

We denote by $\DG^{(\ell)}[\phi,\g]$ and $\DGr[\phi,\g;\de_i^{(\ell)}]$
the ST functional and its relevant part at loop $\ell$.
We prove the following two properties:

\noindent
{\it Locality}: if for any loop
\beq\nome{hyp}
\DGr[\phi,\g;\de^{(\ell)}]=0
\,,
\eeq
then for any loop
\beq\nome{th}
\DG^{(\ell)}[\phi,\g]=0
\,.
\eeq

\noindent
{\it Solvability}:
The set of equations \re{hyp} can be solved perturbatively
by appropriately fixing the couplings $\r_i^{(\ell)}$ at
loop $\ell$ as functions of the couplings $\r_i^{(\ell')}$
at lower loops $\ell'<\ell$.

At zero loop one has $\G^{(\ell=0)}[\phi,\g]=\sbrs[\phi,\g] $
which implies $\DG^{(\ell=0)}[\phi,\g]=0$.
{}From this and the above properties one concludes, by induction on the
number of loops, that the ST identities are satisfied perturbatively.

We shall call the condition \re{hyp}, \ie $\de_i=0$,
the fine tuning equations.
The second property is a consequence of the consistency condition
\beq\nome{cc}
{\cal S}_{\G'}\,\D_\G={\cal S}^2_{\G'}\,\G'=0
\,.
\eeq
To show that \re{cc} ensures the solvability of the fine
tuning equations is relatively simple for $\mu = 0$.
In this case one can check by inspection that the relevant part of
\re{cc} gives enough relations among the parameters $\de_i$ to make
solvable the fine tuning equations (for example in the $SU(2)$ case only
six $\de_i$ are independent).
Here we work at $\mu \ne 0$ and use RG flow method
to prove solvability.
The difficulty arising from the presence of a non-vanishing
subtraction point is due to the fact that by applying $T_6^{(\mu)}$ to
$S_{\G'}\D_\G$, \ie by taking the relevant part, one introduces not
only the relevant part of $\D_\G$ but also some irrelevant vertices
evaluated at the subtraction points. Then one has to study also the
vanishing of irrelevant parts. This will be shown by using the RG flow.

\vskip 0.3cm
\noindent
{\bf{2. Renormalization group flow}}
\vskip 0.3cm
\noindent
In order to prove these properties by using the RG flow
method, we recall first how the effective action
$\G[\phi,\g]$ and the functional $\D_\G[\phi,\g]$ are obtained within
the RG formulation. For more details see \cite{B,BDM}.

($i$) One introduces the Wilsonian effective action
$\se[\phi,\g;\L,\L_0]$ which is obtained by path integration over
the fields with frequencies $\L^2<p^2<\L_0^2$. In this functional one
uses propagators with IR and UV cutoff function $K_{\L\L_0}(p)$ equal
to one in the above region and rapidly vanishing outside.
{}From this definition one has that the coefficients of the monomials
quadratic in the fields in $\se[\phi,\g;\L,\L_0]$ are proportional
to $\hbar$, thus at zero loop order $\se^{(\ell=0)}[\phi,\g;\L,\L_0]$
does not contain quadratic monomials.
This property will play a key r\^ole in the
perturbative proof of locality and solvability by RG flow method.
At the physical point $\L=0$ and \UV the functional $\se[\phi,\g;0,\infty ]$
generates the amputated connected Green functions.
The physical effective action $\G[\phi,\g]$ is then obtained from
$\se[\phi,\g;0,\infty]$ by Legendre transform so that the relevant
parameters $\r_i$ in \re{tG} are simply related to the relevant parameters of
$\se[\phi,\g;0,\infty]$.

{}From the definition of $\se[\phi,\g;\L,\L_0]$ one finds that this
functional satisfies an evolution equation \cite{W,P} in the IR cutoff $\L$
\beq\nome{RG}
\dL \frac{\se}{\hbar} = (2\pi)^8\hbar
\int_p \frac {1}{p^2}\, \dL K_{0\L}(p) \;e^{-i\set/\hbar}\;
\biggl\{\half \frac{\de}{\de A_\mu^a(-p)}\frac{\de}{\de A_\mu^a(p)}
+\frac{\de}{\de c^a(-p)}\frac{\de}{\de\bc^a(p)}
\biggr\}\;e^{i\set/\hbar}\,,
\eeq
where $\int_p\equiv\int\frac{d^4p}{(2\pi)^4}$ and $K_{0\L}(p)$ is the
cutoff function vanishing for $p^2> \L^2$.
This equation can be integrated by giving boundary conditions in $\L$.
As boundary conditions for the RG flow one assumes that at the UV
point $\L=\L_0 \to \infty$ the irrelevant part of the Wilsonian
action vanishes
\beq\nome{locse}
\sei[\phi,\g;\L_0,\L_0] \equiv
(1-T_4^{(\mu)})\, \se[\phi,\g;\L_0,\L_0] \to 0\,,
\;\;\;\;\;\;
\mbox{for}
\;\;\;\;\;\;
\L_0 \to \infty
\,.
\eeq
As far as the point where to fix $\ser[\phi,\g;\L,\L_0]$ is concerned,
one may choose the physical point $\L=0$ and \UV where
$\ser[\phi,\g;0,\infty]$
is given in terms of the couplings $\r_i$ in \re{tG}.
In this formulation the usual loop expansion of the effective action
can be obtained by solving
iteratively the RG evolution equation \re{RG} with these boundary
conditions and the zero loop input $\G^{(\ell=0)}[\phi,\g]=\sbrs[\phi,\g]$.
As shown by Polchinski \cite{P}, the RG formulation provides a
very simple method to prove perturbative renormalizability, \ie
the limit \UV can be taken. The proof can be applied to the gauge
theory case \cite{YM}. RG supplies also a simple method
to prove \cite{YM} that a massless theory is IR finite in perturbation
theory, \ie the limit \IR can be taken, provided a non-vanishing subtraction
point $\mu \ne 0$ is introduced.

($ii$) One then considers the ST identities in \re{ST}. It can be
shown that these identities can be formulated directly for the
Wilsonian action $\se[\phi,\g;\L,\L_0]$ for any $\L$ and $\L_0$ even
away from the physical point $\L=0$ and \UV.
One introduces the generalized BRS transformation
$$
\de A_\mu^a(p)=-\frac i g \eta p_\mu c^a(p) +K_{0\L}(p)\eta\,
\frac{\de \se}{\de u_\mu^a(-p)}\,,
$$
$$
\de c^a(p)=K_{0\L}(p)\eta \,\frac{\de \se}{\de v^a(-p)}\,,
\;\;\;\;\;\;\;\;\;\;\;\;
\de \bc^a(p)=\frac i g \eta p_\mu A_\mu^a(p)\,,
$$
with $\eta$ a Grassmann parameter and $a$ the gauge index.
The cutoff function is cancelled by
an inverse cutoff function entering in the sources.
{}From this transformation one finds the following form of the ST
identities
\beq\nome{STeff}
\De[\phi,\g;\L,\L_0]=0
\eeq
where
$$
\De[\phi,\g;\L,\L_0] \equiv
\int_p K_{0\L}(p) e^{-i\set}\,
\biggl\{
\frac{\de}{\de A^a_\mu(p)}\frac{\de}{\de u^a_\mu(-p)} +
\frac{\de}{\de c^a(p)}\frac{\de}{\de v^a(-p)}
\biggr\}\,e^{i\set}
$$
$$
+ i \int_p \biggl\{ p^2 A^a_\mu(p)\frac{\de}{\de u^a_\mu(p)}
+\frac i g p_\mu c^a(p)\frac{\de}{\de A^a_\mu(p)}
+\frac i g p_\mu w^a_\mu(p)
\frac{\de}{\de v^a(p)}
-\frac i g p_\mu A^a_\mu(p)\frac{\de}{\de \bc^a(p)} \biggr\} \se
\,.
$$
For $\L=\L_0 \to \infty$ we have $K_{0 \L_0}(p) \to 1$ and
$\se$ becomes local (see \re{locse}). Therefore in this limit we have
\beq\nome{Dirr}
\Dei[\phi,\g;\L_0,\L_0] \equiv
(1-T_5^{(\mu)})\,\De[\phi,\g;\L_0,\L_0] \to 0\,,
\;\;\;\;\;\;
\mbox{as }
\;\;\;\;\;\;
\L_0 \to \infty
\,.
\eeq
At the physical point $\L=0$ and \UV,
$\De[\phi,\g;0,\infty]$ is related to $\DG[\phi,\g]$ via Legendre
transform, \ie the vertices of $\DG[\phi,\g]$
are obtained from the vertices of $\De[\phi,\g;0,\infty]$
by neglecting the one particle ($A,c$) reducible contributions.

($iii$)
The RG flow for the functional $\De$ is given by the following linear
evolution equation
\beq\nome{Dflow}
\dL \De[\phi,\g;\L,\L_0]= \{M_1[\se]+\hbar M_2\}\cdot
\De [\phi,\g;\L,\L_0]\,,
\eeq
where $M_1$ and $M_2$ are
$$
M_1[\se]=-(2\pi)^8\int_p \frac{1}{p^2} \dL K_{0\L}(p)
\biggl\{\frac{\de\se}{\de A^a_\mu(p)}
\frac{\de}{\de A^a_\mu(-p)}
+\frac{\de\se}{\de c^a(p)}
\frac{\de}{\de \bc^a(-p)}\;\; -\;\; c\leftrightarrow \bc\;\;
\biggr\}\,,
$$
$$
M_2=\frac i 2 (2\pi)^8\int_p \frac{1}{p^2} \dL K_{0\L}(p)
\biggl\{
\frac{\de}{\de A^a_\mu(p)}\frac{\de}{\de A^a_\mu(-p)}
+ \frac{\de}{\de c^a(p)}
\frac{\de}{\de \bc^a(-p)}\;\; -\;\; c\leftrightarrow \bc\;\;
\biggr\}\,.
$$
The boundary conditions for \re{Dflow} are obtained from the ones of
$\se[\phi,\g;\L,\L_0]$.
At the UV point the irrelevant part vanishes (see \re{Dirr}) and
at the physical point $\L=0$ and \UV the relevant part is given
by the parameters $\de_i$.
By taking advantage of \re{Dirr} and perturbative renormalizability
in the following we will consider the limit \UV.

\vskip 0.3cm
\noindent
{\bf{3. Locality}}
\vskip 0.3cm
\noindent
We now prove the property of locality \re{hyp}, \re{th}. To do this
we show that from the hypothesis \re{hyp} we have
\beq\nome{th'}
\De^{(\ell)}[\phi,\g;\L,\infty]=0
\eeq
for any loop and any value of $\L$ and this implies the identity
\re{th}. Indeed at the physical point $\L=0$ the condition
$\De^{(\ell)}[\phi,\g;0,\infty ]=0$ becomes, via Legendre transform,
the ST identities $\DG^{(\ell)}[\phi,\g]=0$.

The proof is obtained by induction on the number
of loops and on the number of fields $n$ in the vertices of
$\De[\phi,\g;\L,\infty]$.
A crucial point in the inductive proof is the fact that if
$\De^{(\ell')}[\phi,\g;\L,\infty]=0$ for any $\ell'<\ell$,
then at loop $\ell$, the RG flow \re{Dflow} becomes
\beq\nome{Dflow'}
\dL \De^{(\ell)}[\phi,\g;\L,\infty]=
M_1[\se^{(\ell=0)}]\cdot\De^{(\ell)} [\phi,\g;\L,\infty]
\,,
\eeq
where $M_1[\se^{(\ell=0)}]$ does not contain monomials linear in the
fields or sources (as observed before $\se^{(\ell=0)}[\phi,\g]$ does
not contain quadratic monomials).

First we consider the case $\ell=0$ and prove that
$\De^{(\ell=0)}[\phi,\g;\L,\infty]=0$ by induction on $n$.
We denote by $\Den^{(\ell)}(\cdots;\L,\infty)$ the vertices of $\De$
at loop $\ell$ with $n$ fields or sources.
The dots denote momenta, internal and Lorentz indices.
For $n=2$, \ie for the coefficient of the $Ac$ monomial,
$\Dee^{(\ell=0)}(p;\L,\infty)$, we have from \re{Dflow'}
$$
\dL \Dee^{(\ell=0)}(p;\L,\infty)=0
\,.
$$
By using the boundary condition \re{Dirr} we conclude that
$\Dee^{(\ell=0)}(p;\L,\infty)$ is given by its relevant part at the
physical point.
Since in $\Dee$ there are no one particle reducible contributions,
this vertex at the physical point is the same as the
$Ac$-vertex of $\DG$ which, according to the hypothesis \re{hyp},
has no relevant part.
Therefore also $\Dee$ has no relevant part and we have that the full
vertex vanishes, $\Dee^{(\ell=0)}(p;\L,\infty)=0$.

Now we proceed by iteration on $n$, namely we suppose that
$\Denn^{(\ell=0)}(\cdots;\L,\infty)=0$ for all $n'<n$ and show that
$\Den^{(\ell=0)}(\cdots;\L,\infty)=0$.
Using again the fact that $\se^{(0)}$ has no quadratic monomials,
we have from \re{Dflow'}
$$
\dL \Den^{(\ell=0)}(\cdots;\L,\infty)=0
\,.
$$
By using the boundary condition
\re{Dirr}, we have that $\Den^{(\ell=0)}(\cdots;\L,\infty)$ is
given by its relevant part at the physical point,
which is zero as a consequence of the hypothesis
\re{hyp}. This is due to the fact that the $n$-vertices of $\De$
and $\DG$ are the same.
Indeed the one particle reducible contributions of the $n$-vertices
of $\De$ involve vertices with $n'<n$ which are zero because of the
inductive hypothesis. Therefore we have in general
$\De^{(\ell=0)}[\phi,\g;\L,\infty]=0$.

Consider now the case $\ell>0$. We assume
$\De^{(\ell')}[\phi,\g;\L,\infty]=0$ for any $\ell'<\ell$ and
we want to prove \re{th'} at loop $\ell$.
The proof consists in repeating exactly the same procedure of induction
on $n$ which has been followed above for the zero loop case.
This concludes the proof that $\De^{(\ell)}[\phi,\g;\L,\infty]=0$
for any $\ell$.

\vskip 0.3cm
\noindent
{\bf{4. Solvability at $\mu \ne 0$}}
\vskip 0.3cm
\noindent
We show here that the fine tuning equations \re{hyp} can be solved for
$\mu \ne 0$ as a consequence of the consistency condition \re{cc}.
Namely for $SU(2)$ one can fix the six couplings $\r_i$ of
the effective action \re{tG} in such a way that the eleven relevant
parameters $\de_i$ vanish.
For $\mu =0$ the proof of the solvability of \re{hyp} is a direct
consequence of the consistency condition \re{cc} which gives \cite{B}
$$
g\de_2=\de_3+\de_{4}\,,
\;\;\;\;\;\;\;\;\;
\de_{6}=-g\de_5\,,
\;\;\;\;\;\;\;\;\;
\de_7=\de_9=\de_{11}\,,
\;\;\;\;\;\;\;\;\;
2\de_8=\de_{10}\,.
$$
The general relevant functional $\DG$ which satisfies the above
relations can be expressed as the BRS variation of a local functional,
namely
\beq\nome{repr}
\DGr[\phi,\g;\de_i]=
{\cal S}_{{\G'}^{(0)}}\,
\Gr[\phi,\g;{\bar \r}_i]
\,.
\eeq
This expresses the triviality of the ghost number one cohomology class
of the Slavnov operator for the $SU(2)$ Yang-Mills theory.
If this representation holds, then the eleven parameters $\de_i$ are
the following functions of the six parameters $\bar \r_i$
\beq\nome{bard}
\eqalign{
&
\de_1=\frac 1 g \br_1
\,,\;\;\;\;\;\;
\de_2=\frac 1 g \br_2
\,,\;\;\;\;\;\;
\de_3=\br_5
\,,\;\;\;\;\;\;
\de_4= g \de_2 -\de_3=\br_2-\br_5\,,
\cr&
\de_6=- g \de_5= \br_5-\br_6
\,,\;\;\;\;\;\;
\de_7=\de_9=\de_{11}=g (\br_3-\br_4-\br_5)
\,,\;\;\;\;\;\;
2\de_8=\de_{10}=2g (\br_5-\br_3)
\,.
}
\eeq
For $\mu \ne 0$ the representation \re{repr} is not valid a priori,
since the consistency condition \re{cc} gives relations among the
$\de_i$ which involve also irrelevant contributions of the $\DG$
vertices evaluated at some subtraction point $\mu \ne 0$.
However we have seen that the irrelevant parts can be set to zero
iteratively. Before showing how this procedure works we recall how,
even in the case of a non-vanishing subtraction point $\mu \ne 0$,
the fine tuning equations can be solved loopwise if the
representation \re{repr} holds.
{}From \re{ST} we have
$$
\DG^{(\ell)}[\phi,\g]
=2\,{\cal S}_{{\G'}^{(0)}}\,{\G'}^{(\ell)}
+\sum_{k=1}^{\ell-1} {\cal S}_{{\G'}^{(k)}}\,{\G'}^{(\ell-k)}\,.
$$
By applying $T^{(\mu)}_5$, we obtain the relevant part
\beq\nome{Drell}
\DGr[\phi,\g;\de_i^{(\ell)}]
=
2\,{\cal S}_{{\G'}^{(0)}}\,{\Gr'}[\phi,\g;\r_i^{(\ell)}]
+\Omega^{(\ell)}[\phi,\g]\,,
\eeq
where
$$
\Omega^{(\ell)}[\phi,\g]=T^{(\mu)}_5\,\sum_{k=1}^{\ell-1}
{\cal S}_{{\G'}^{(k)}}\,{\G'}^{(\ell-k)}
+2\, \left(
T^{(\mu)}_5\, {\cal S}_{{\G'}^{(0)}}
-{\cal S}_{{\G'}^{(0)}}\,T^{(\mu)}_4
\right)\;
{\G'}^{(\ell)}\,.
$$
The crucial observation now is that $\Omega^{(\ell)}$
depends only on the relevant parameters $\r_i^{(\ell')}$ at lower
loops $\ell'<\ell$.
This is obvious, since the product of two relevant vertices is a
relevant vertex so that
$$
T^{(\mu)}_5\, {\cal S}_{{\G'}^{(0)}}\, T^{(\mu)}_4
= {\cal S}_{{\G'}^{(0)}}\, T^{(\mu)}_4
\,.
$$
As a consequence $(T^{(\mu)}_5\, {\cal S}_{{\G'}^{(0)}}
-{\cal S}_{{\G'}^{(0)}}\,T^{(\mu)}_4)\,{\Gr'}=0$ and
$\Omega^{(\ell)}$ does not receive contribution
from the couplings $\r_i^{(\ell)}$.
{}From eqs.~\re{repr} and \re{Drell} one has that $\Omega^{(\ell)}$
must be of the form
$$
\Omega^{(\ell)}=
{\cal S}_{{\G'}^{(0)}}\,
\Gr'[\phi,\g;{\r'}_i^{(\ell)}]\,,
\;\;\;\;\;\;
\;\;\;\;\;\;
{\r'}_i^{(\ell)}=\bar \r_i^{(\ell)}-2\r_i^{(\ell)}
\,,
$$
where ${\r'}_i^{(\ell)}$ are given in terms of
the couplings $\r_i^{(\ell')}$ at lower loops $\ell' < \ell$.
Therefore one has $\bar \r_i=0$, \ie $\D^{(\ell)}_{\G}[\phi,\g]=0$
if one sets
\beq\nome{sol}
\r_i^{(\ell)}=-\frac 1 2 {\r'}_i^{(\ell)}
\,.
\eeq
This ends the proof of the fact that if \re{repr} holds then the
fine tuning equations can be solved even for $\mu\ne 0$.

We now come to discuss whether the representation \re{repr} can be
used also for $\mu \ne 0$.
As recalled before \re{repr} is not valid if there are
irrelevant contributions in the various vertices of $\DG$.
To show how to use \re{repr} in this case one proceeds as follows by
exploiting the perturbative RG results obtained in the previous
section.

($i$) The $Ac$-vertex $\D_{\G,2}$ is given only by its relevant part
(see proof of locality), \ie given by the parameters
$\de_1$ and $\de_2$ in \re{DGrel}.
Therefore for this vertex the representation \re{repr} holds
and from \re{sol} one can fix $\r_1$ and $\r_2$ in such a way
the full vertex $\D_{\G,2}$ vanishes.
One shows \cite{BDM} that the equations $\de_1=\de_2=0$ are
solved by $\r_1=\r_2=0$.

($ii$) Once $\D_{\G,2}=0$, the $AAc-$ and $wcc-$vertices
$\D_{\G,3}$ are given only by their relevant parts
(see proof of locality), \ie given by the parameters
$\de_3$, $\de_4$ and $\de_5$ in \re{DGrel}.
Therefore, for these relevant vertices the consistency condition
\re{cc} gives
$$
\de_3+\de_{4}=0\,.
$$
Then the representation \re{repr} holds and one can fix the couplings
$\r_3,\,\r_4$ in such a way that $\D_{\G,3}=0$.
One shows that the equations $\de_3=\de_4=\de_5=0$ are solved by
$\r_3=0$ and $\r_4$ given by an irrelevant part of the $wcA$-vertex of
$\G[\phi,\g]$ evaluated at some Euclidean subtraction point.
For the exact evaluation of $\r_4$ in $SU(2)$ see ref.~\cite{BDM}.

($iii$) Once $\D_{\G,3}=0$, the $AAAc-$ and $wccA-$vertices
$\D_{\G,4}$ are given only by their relevant parts
(see again proof of locality), \ie given by the parameters
$\de_6,\ldots,\de_{11}$ in \re{DGrel}.
For these relevant vertices the consistency condition gives
$$
\de_{6}=0\,,
\;\;\;\;\;\;\;\;\;
\de_7=\de_9=\de_{11}\,,
\;\;\;\;\;\;\;\;\;
2\de_8=\de_{10}\,,
$$
so that the representation \re{repr} holds and one can fix $\r_5$ and
$\r_6$ in such a way $\D_{\G,4}=0$.
One shows that the equations $\de_6= \dots =\de_{11}=0$ are solved
by $\r_5$ and $\r_6$ given by irrelevant parts of
$wcA-$, $wcAA-$, $AAA-$ and $AAAA-$vertices of $\G[\phi,\g]$ evaluated
at some Euclidean subtraction points.
For the exact evaluation of $\r_5$ and $\r_6$ in $SU(2)$ see ref.~\cite{BDM}.

\vskip 0.3cm
\noindent
{\bf{5. Comments}}
\vskip 0.3cm
\noindent
By using the exact RG flow we have proved perturbatively
that the ST identities \re{ST} are valid provided
that the relevant part of the effective action is properly chosen.
In the $SU(2)$ case one has
$$
\Gr[\phi,\g;\r_i] =\sbrs+
\hbar
\int d^4x \biggl\{
\frac {\r_4}{2} \, v\cdot c \wedge c
+\frac{\r_5 g^2}{4}(A_\mu\wedge A_\nu)^2
+\frac{\r_6 g^2}{4} (A_\mu\cdot A_\nu)^2
\biggr\},
$$
where the only non-vanishing couplings $\r_4$, $\r_5$ and $\r_6$
are given in terms of appropriate irrelevant vertices of $\G$
evaluated at the subtraction point \cite{BDM}.
This form allows one to perform the perturbative
expansion since irrelevant vertices at loop $\ell$ involve
relevant couplings at lower loops $\ell'<\ell$.

The method is general. As shown in ref.~\cite{YM} it can be applied for
instance to $SU(2)$ gauge theory with fermions.
The application to the case of chiral gauge theories without anomalies
should be also possible along the same lines.

The RG method provides in principle a non-perturbative formulation
thus it could be used to extend the proof of locality
and solvability beyond perturbation theory.
It is then important to pin down the points where perturbation theory
was needed in this work.
Here the proof of locality is essentially based on the following two facts:
1) in the RG flow \re{Dflow} for $\De[\phi,\g;\L,\infty]$ we used only
the linear operator $M_1$ and neglected $M_2$.
This is possible only in perturbation theory because
of the inductive hypothesis;
2) we used the fact that $M_1$ does not contain monomials linear in
the fields, which is true only if one uses, via induction, the tree
approximation of $\se$.
For the proof of solvability one uses the locality of $\D_\G$ (see
second part of sect.~4) to set to zero irrelevant contributions in the
consistency condition \re{cc}. Due to this fact also the proof of
this property is restricted to the perturbative framework.
Thus, it seems that the crucial point is the non-perturbative
extension of locality.

\vspace{3mm}\noindent
We have benefited greatly from discussions with C. Becchi and M.
Tonin.

\end{document}